\documentclass[11pt,twocolumn]{article}
\usepackage{amsmath}
\usepackage{amsfonts}
\usepackage{graphicx}
\usepackage[super,articletitle=true]{achemso}
\usepackage{hyperref}
\usepackage{geometry}
\usepackage{titlesec}
\usepackage{fancyhdr}
\usepackage{xcolor}
\usepackage{subcaption}
\usepackage[font=small]{caption}
\makeatletter
\renewcommand{\thefootnote}
\makeatother
\titleformat{\section}{\large\bfseries}{\thesection.}{1em}{}
\titleformat{\subsection}{\normalsize\bfseries}{\thesubsection.}{1em}{}
\title{\Large \textbf{Local Structure and Dynamics of Three-Dimensional Covalent Organic Frameworks}}
\author{
    \begin{minipage}[t]{\textwidth}
    \centering
{Francesco Tavani}\textsuperscript{1,2,3,*}, Saber Mirzaei\textsuperscript{1}, Jian Yin\textsuperscript{1}, Yen-hsu Lin\textsuperscript{1}, Caden Myers\textsuperscript{4}, Cheng-Hung Lin\textsuperscript{5}, A. M. Milinda Abeykoon\textsuperscript{5},
Simon J. L. Billinge\textsuperscript{6}, Omar M. Yaghi\textsuperscript{1,7,*}\\
    \vspace{0.5cm}
    \textsuperscript{1}Department of Chemistry, University of California, Berkeley, California 94720, USA \\
    \textsuperscript{2}Bakar Institute of Digital Materials for the Planet, Division of Computing, Data Science, and Society, University of California, Berkeley, California 94720, USA \\
    \textsuperscript{3}Dipartimento di Chimica, Università degli Studi di Roma La Sapienza,\\ P.le A. Moro 5, I-00185, Rome, Italy \\
    \textsuperscript{4}Department of Applied Physics and Applied Mathematics, Columbia University, New York, New York 10027, USA\\
    \textsuperscript{5}National Synchrotron Light Source II, Brookhaven National Laboratory, Upton, \\New York 11973, USA\\
    \textsuperscript{6}Department of Materials, University of California Santa Barbara, Santa Barbara, California 93106, USA\\ 
    \textsuperscript{7}Current address: Departments of Chemistry and Chemical Engineering, and AIMATRY, Tsinghua University, Beijing, 100084, China\\  
    \vspace{0.5cm}
    \textsuperscript{*}
    Corresponding authors: ftavani@berkeley.edu; yaghi@tsinghua.edu.cn
    \end{minipage}}
    
\date{}

\begin{document}
\twocolumn[
\maketitle
\begin{center}
    \large \textbf{Abstract}
\end{center}
\begin{center}
    \begin{minipage}{0.9\textwidth}
        \normalsize

Resolving and controlling the local dynamical properties of covalent organic frameworks (COFs) remains a central challenge, particularly when assembled from large, flexible building units. Here, we combine synchrotron X-ray pair distribution function (PDF) analyses with machine learning-accelerated molecular dynamics (MD) simulations to resolve the local structure and dynamics of two three-dimensional imine-linked COFs, COF-682 {[(DHP)(TAM)]$_{imine}$}, assembled from 6,13-dihydropentacene (DHP) and tetrakis(4-aminophenyl)methane (TAM), and COF-612 {[(HBC-LA$_{12}$)(HAPT)$_2$]$_{imine}$}, assembled from nanographene dodecabenzaldehyde hexakis{[3,5-bis($p$-formylphenyl)-4,6-dimethoxyphenyl]}hexabenzocoronene (HBC-LA$_{12}$) and  2,3,6,7,14,15-hexa(4-aminophenyl)triptycene (HAPT). Validated against the experimental PDFs through ensemble-averaged calculations, the simulations show that the exposed $\pi$-surface and V-shaped geometry of the DHP linker endow COF-682 with enhanced local flexibility through face-to-face and offset $\pi$-stacking interactions differing in both their average interplanar separation and their ring-plane tilt angle. In contrast, the extended nanographene linker rigidifies COF-612 by maintaining the planarity of its fused cores, while the linker pendant aryl rings equip both COFs with enhanced librational ability. The simulations further provide quantitative measures of the translational and reorientational mobility of the linkers, revealing how local COF dynamics may be tuned by balancing non-covalent interactions and different degrees of aromatic rigidity. The PDF-MD experimental-computational approach holds promise as a general method beyond conventional crystallography to gain insights into the local properties of COFs with the aim of directing their dynamic function.

  \end{minipage}
\end{center}
]

\clearpage

\section*{Introduction}
\begin{figure*}[h!]
    \centering
    \includegraphics[width=\linewidth]{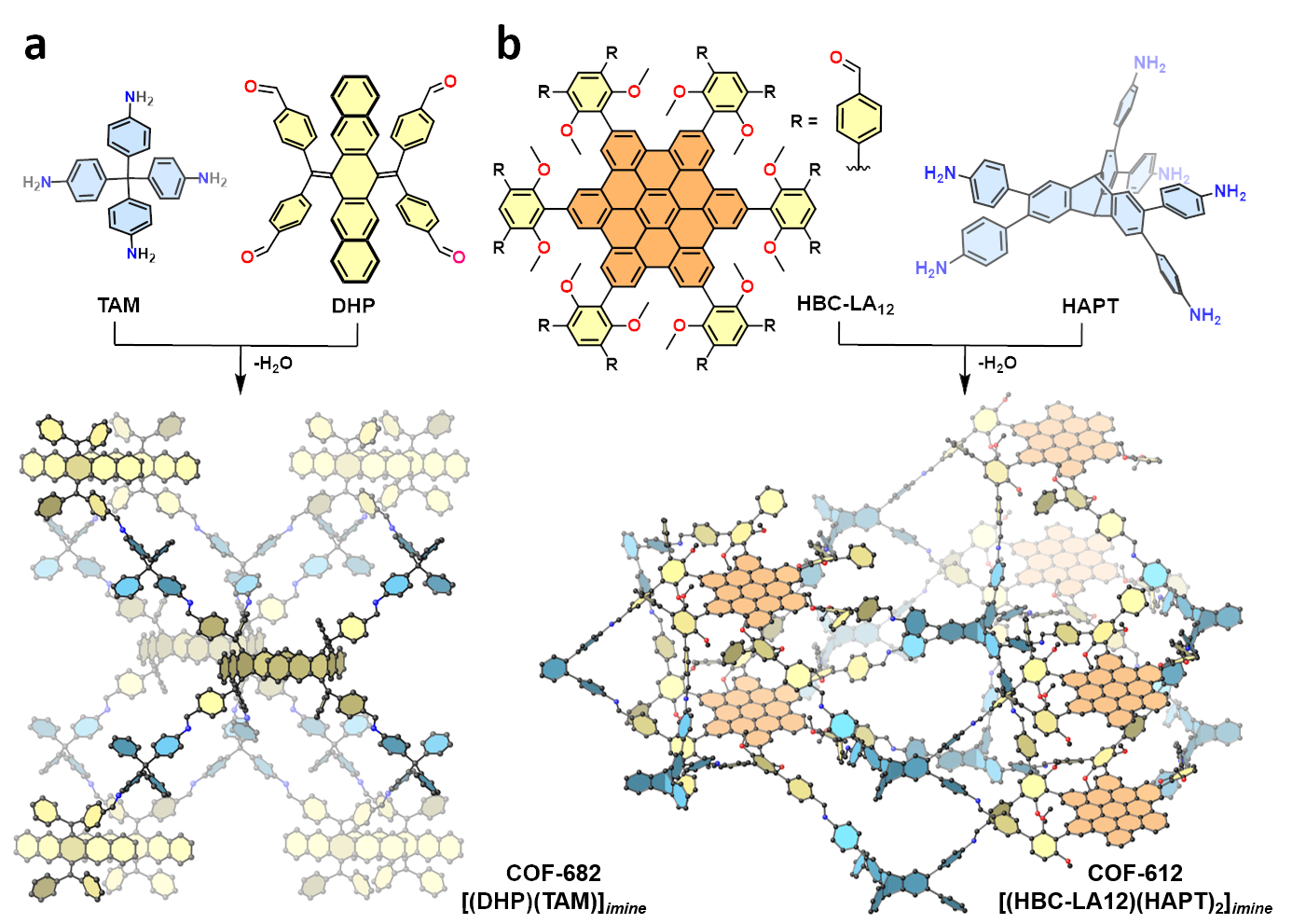}
\caption{Synthesis of COF-682 and COF-612. (a) Condensation of the V-shaped tetratopic aldehyde DHP with the tetrahedral tetraamine TAM to obtain COF-682. (b) Condensation of the twelve-connected, dodeca-benzaldehyde-functionalized
nanographene HBC-LA$_{12}$ with the six-connected, trigonal-prismatic hexaamine HAPT to obtain COF-612.}
\label{fig:synthesis}
\end{figure*}

Covalent organic frameworks (COFs) are porous materials in which organic molecular building units are joined by covalent bonds into predesigned, tunable structures that exhibit
high thermal and mechanical stability.\cite{Cote,COF2,COF_account,diercks,MOF3_f} COFs hold promise
for gas storage,\cite{cof_co2_n1,COF_water_n2} separation,\cite{COF_sep} and catalysis,\cite{Guo}
among other applications, where performance relies on accurate knowledge of the framework structure
and its real-time evolution under operating conditions. The strong covalent bonds that
underlie their stability, however, make COF crystallization difficult\cite{Haase_COF,COFS_cryst_zhang_jacs}
and often limit the size and quality of the resulting crystals, posing a significant challenge to
molecular-level analysis of COF structure and dynamics.\cite{daliran_adv_funct_mat,chem_mat} To date,
the structural characterization of COFs has relied mainly on powder X-ray diffraction (PXRD)\cite{Cote}
and, where crystals allow, single-crystal\cite{tianqiong,chen_chem_soc_rev} or electron
diffraction,\cite{COFS_cryst_zhang_jacs,chen_chem_soc_rev} complemented by spectroscopic probes such
as Fourier-transform infrared and solid-state NMR spectroscopy, used mainly to confirm linkage
chemistry.\cite{frimpong_ftir,MOF3_f,wang_nmr_jacs} Indeed, the COF frameworks are composed entirely of light elements (C, N, O, B, H) and offer no absorbing center to hard X-ray spectroscopic probes, such as hard X-ray absorption spectroscopy (XAS) and X-ray emission spectroscopy (XES).\cite{tav_dalton,tav_pccp,delgiudice} 
However, PXRD is of limited diagnostic value for less ordered frameworks, since the light-element composition and high porosity of COFs limit the resolution of their PXRD profiles and Bragg intensities are attenuated, broadened, or extinguished as disorder increases,\cite{billings_pdf_review} while spectroscopy reports local bonding and electronic properties but is insensitive to long-range length scales.\cite{delgiudice,tav_dalton,tav_pccp,catal13101338} As a result, neither approach fully captures how framework atoms are arranged, or how they move, beyond the ideal periodic lattice.\cite{daliranreview} Yet it
is precisely the local arrangement of atoms and the flexibility of the framework---linker libration,
ring rotation, and translational freedom---that govern adsorption, diffusion, and mechanical
response in reticular materials.\cite{nanoporous_flexibility,CCS} Resolving these local properties, ideally \emph{in situ} on a material in its working state, remains a longstanding challenge.\\
The atomic pair distribution function (PDF) method is well suited to meet this
challenge.\cite{terban_chem_rev,billings_pdf_review} In contrast to conventional diffraction, a PDF
experiment analyzes the Bragg and diffuse scattering simultaneously to yield a real-space histogram,
$G(r)$, of interatomic distances, providing direct access to short- and intermediate-range
order in crystalline, defective, and amorphous solids alike.\cite{terban_chem_rev} Although these
attributes have established PDF as a powerful probe of materials on the nanoscale,\cite{Griffiths2024_NatMater,Klove2023_AdvMater,goodwin_n1,Zhu2021_AdvSci}
its application to reticular systems such as COFs has so far been largely confined to tracking structural
transitions\cite{PlateroPrats2016_JACS} and disorder in metal--organic frameworks
(MOFs).\cite{Cliffe2014_NatCommun,RomeroMuniz2024_CSR,Bennett2014_AccChemRes,Bennett2010_PRL,Cao2012_ChemCommun,Gaillac2017_NatMater,Keen2018_PCCP,Terban2018_Nanoscale,Molina2021_MicroMeso,Sapnik2022_NatCommun,Sapnik2023_CommunChem}
Its use to study COFs, by contrast, remains limited, with only a handful of reports to
date.\cite{lotsch_chemsci,RomeroMuniz2021_ACSAMI} A central reason is that the structural complexity
of reticular frameworks, and of COFs in particular, makes the quantitative assignment of individual
$G(r)$ peaks to explicit theoretical models difficult, so that the molecular-level information encoded
in the PDF has remained largely out of reach.\\
Here, we combine synchrotron X-ray PDF analysis with molecular dynamics (MD) simulations based on machine learning interatomic potentials (MLIPs) to resolve
the local structure and dynamics of two three-dimensional imine-linked COFs, COF-682 \cite{MOF3_f} and COF-612 \cite{preprint_saber}, deliberately chosen to differ
in the geometry and rigidity of their building units. COF-682 is obtained by condensing the V-shaped
tetratopic aldehyde 6,13-dihydropentacene (DHP), whose exposed acene $\pi$-surface bears four pendant 4-formylphenyl arms, with
the tetrahedral tetraamine tetrakis(4-aminophenyl)methane (TAM), yielding the (4,4)-connected framework [(DHP)(TAM)]$_{imine}$ (Figure \ref{fig:synthesis}a).\cite{preprint_saber}
COF-612 is formed by reticulating the twelve-connected, dodeca-benzaldehyde-functionalized nanographene
hexakis[3,5-bis($p$-formylphenyl)-4,6-dimethoxyphenyl]hexabenzocoronene (HBC-LA$_{12}$) with the
six-connected, trigonal-prismatic hexaamine 2,3,6,7,14,15-hexa(4-aminophenyl)triptycene (HAPT), yielding
COF-612, [(HBC-LA$_{12}$)(HAPT)$_2$]$_{imine}$ (Figure \ref{fig:synthesis}b).\cite{MOF3_f} Validating the MD models of COF-682 and COF-612 against the experimental PDFs through
ensemble-averaged PDF calculations, and then interrogating the MD trajectories, we resolve how the
building units of each COF dictate its local order and motion, obtaining quantitative measures
of the translational freedom of every linker, the plane-tilt reorientation of their aromatic rings,
and the positional probability density of all framework atoms. In doing
so, we establish the combined PDF--MD approach as a general method for simultaneously probing the local structure and dynamics of COFs, with the potential to extend from \emph{ex situ} to \emph{in situ} conditions and guide the rational design of COFs with local flexibility.
\section*{Results and Discussion}

The two three-dimensional COFs studied here were constructed by imine condensation (Figures S1--S4, full characterization of samples may be found in \cite{preprint_saber,MOF3_f}). COF-682 crystallizes in the orthorhombic
space group \textit{Pnnm} with \textbf{pts} topology as a non-interpenetrated lattice in which the
DHP cores adopt a uniform, parallel-displaced $\pi$-stacking along the $c$ axis. In particular, the $\pi$-stacking non-covalent
interaction likely directs COF crystallization while the acene core retains its characteristic V-shape.\cite{preprint_saber}
COF-612 adopts the (3,6,6)-connected \textbf{kez}
topology in the hexagonal space group $P\overline{6}c2$ \cite{MOF3_f}. 

\begin{figure*}[h!]
    \centering
    \includegraphics[width=\linewidth]{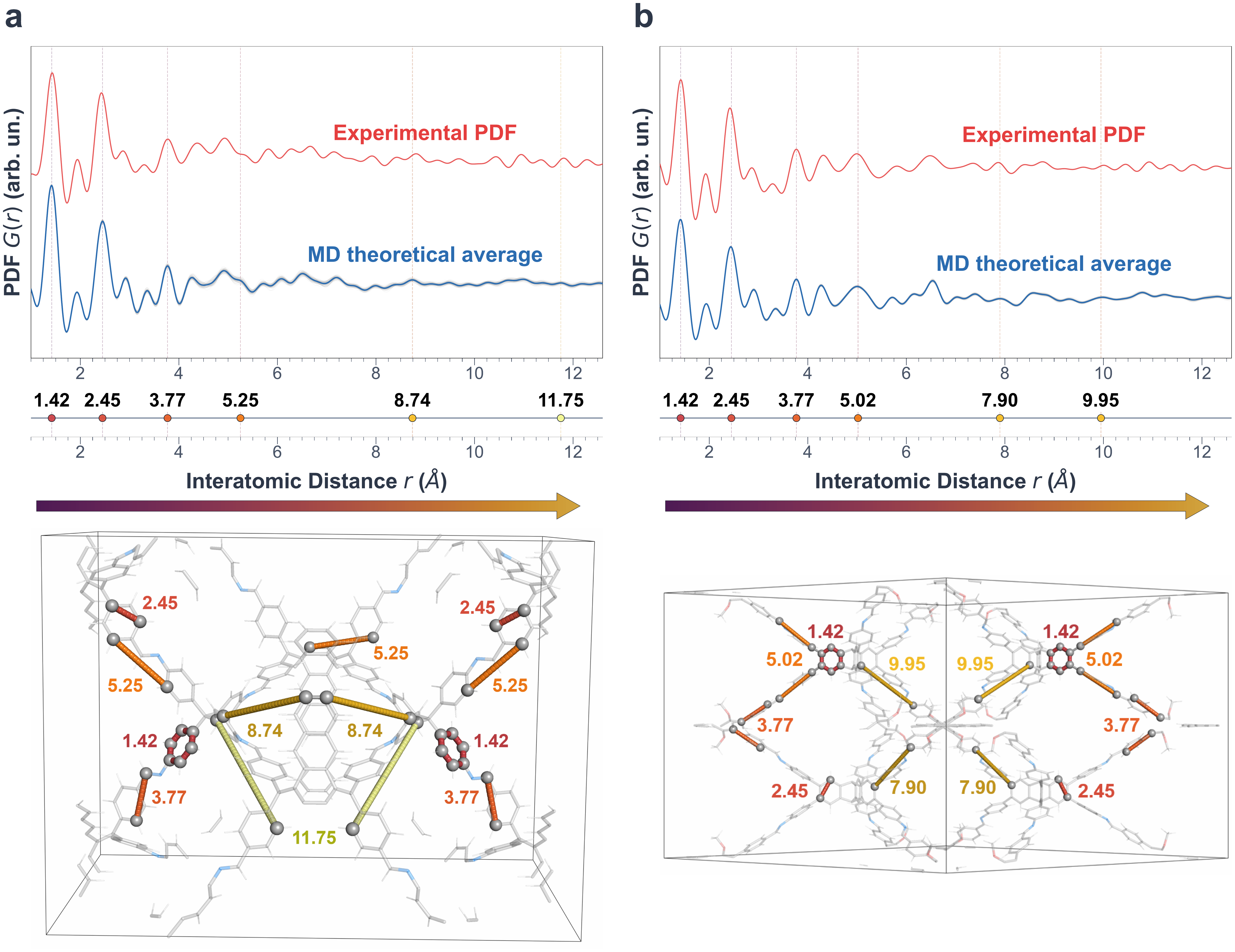}
\caption{Local structure of COF-682 (a) and COF-612 (b) probed by X-ray PDF. \textit{Top}: experimental (red) and MD ensemble-averaged PDFs (blue). Theoretical PDFs calculated from individual MD snapshots are displayed in gray beside the given average PDF curves. Positions of representative PDF peaks are indicated on the distance axes below the top panels. \textit{Bottom}: structural representations of the unit cells of COF-682 and COF-612, where selected intra- and inter-linker distances matching the previously selected PDF distances are highlighted. Color code: C, dark gray; N, blue; H, light gray.}
\label{fig:PDF}
\end{figure*}

\noindent To gain atomic-scale insights into the local properties of the COFs, we performed a PDF analysis of total X-ray scattering data collected on both COF samples at 290~K using synchrotron radiation ($\lambda = 0.1665$~\AA{}) \cite{Juhas:nb5046,xpdfsuite}. The resulting PDFs of COF-682 and COF-612, shown in Figures~\ref{fig:PDF}a,b (red curves), display a series of well-defined peaks that persist above $\sim$12~\AA{}, indicating that both frameworks retain well-defined short- and intermediate-range order. The PDFs can be divided into three regions: (i) a low-$r$ region (below $\sim$3~\AA{}) dominated by intramolecular bonds and intra-ring correlations that are essentially identical for the two frameworks, reflecting the common phenylene and fused-ring units from which both COFs are built; (ii) an intermediate-$r$ region ($\sim$3--7~\AA{}) governed by intra-linker vectors that span the fused cores and the pendant aryl arms; and (iii) a higher-$r$ region ($>$7~\AA{}) encoding the inter-linker distances that bridge the imine linkage between adjacent building units and the distinct connectivity and linker geometry of the two COF topologies. \\

\noindent To interpret these correlations, we computed the corresponding PDFs using the DiffPy-CMI program \cite{diff_cmi} as the ensemble average over MD simulations\cite{MOF1f,MOF2f} of each COF at 290~K (Figures~\ref{fig:PDF}a,b, blue curves). The MD simulations used the universal model for atoms (UMA),\cite{wood2026umafamilyuniversalmodels} a transferable machine learning interatomic potential (MLIP) trained on large and chemically diverse quantum-mechanical datasets including periodic materials based on organics. In particular, UMA delivers near-DFT accuracy for energies and forces and enables the exploration of length- and time-scales inaccessible to \emph{ab initio} methods (see Section~S2 of the SI for additional details on the employed MLIP-MD approach)\cite{wood2026umafamilyuniversalmodels,Hjorth_Larsen_2017,moritz2016linearlyconvergentstochasticlbfgsalgorithm}. The PDFs computed from the individual MD snapshots are shown as thin gray curves in Figures~\ref{fig:PDF}a,b, while their well-converged ensemble averages (see Figure S5 for convergence tests) are displayed in blue. 
As shown in Figure~\ref{fig:PDF}, the MD-averaged PDFs reproduce the experimental PDFs of both COFs in the position and relative intensity of all main peaks, from the intramolecular bonds at low $r$ to the inter-linker correlations beyond 7~\AA{}. This very good level of agreement validates the MD-based representation of the local and intermediate-range structure of both frameworks. Notably, the snapshot PDFs of COF-682 differ appreciably from one another (Figure~\ref{fig:PDF}a, gray lines), reflecting the conformational flexibility of the framework, whereas those of COF-612 are nearly superimposable, consistent with the rigidity imparted by its nanographene core. 

\begin{figure*}[h!]
    \centering
    \includegraphics[width=0.77\linewidth]{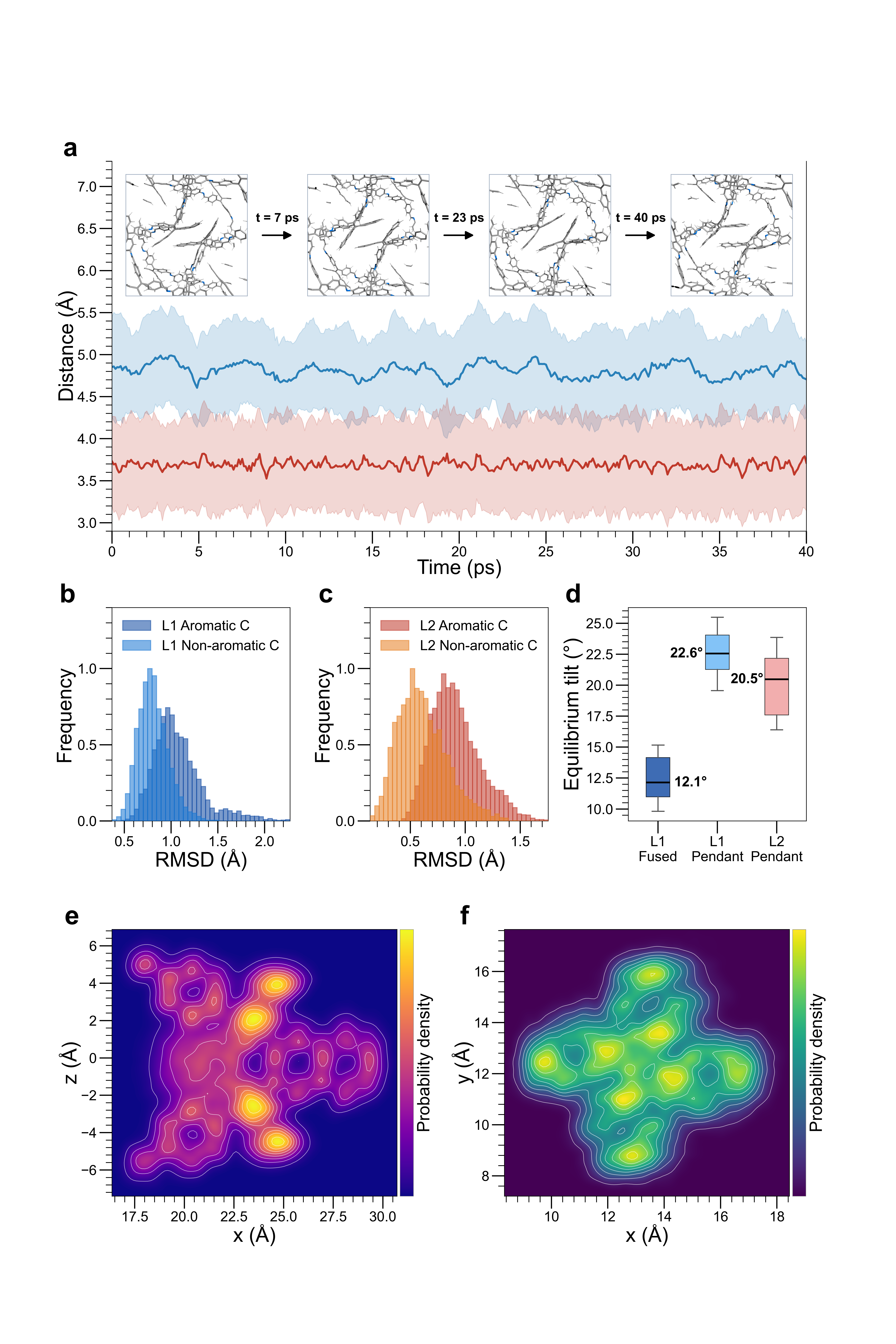}
\caption{MD analysis of the dynamical properties of COF-682. (a) Evolution of the DHP$\cdots$DHP interplanar $\pi$-stacking distance for the face-to-face (blue) and tilted (red) stacking modes, where the solid lines and shaded bands are the MD-averaged distances and the corresponding standard deviations, respectively. The insets show representative MD snapshots
of the framework at $t = 0$, 7, 23, and 40~ps. (b,c) RMSD distribution of the aromatic and non-aromatic carbons of linker L1 (DHP, b) and linker L2 (TAM, c). (d) Box plot distributions of the equilibrium tilt angles of the fused core (L1 fused) and of the pendant aryl rings (L1 pendant, L2 pendant). (e,f) Time-averaged
two-dimensional positional probability densities of representative L1 (e) and L2 linkers (f).}
\label{fig:fig3}
\end{figure*}

\noindent We next used the MD results to assign the individual PDF peaks to specific interatomic vectors, mapping them onto the respective COF unit cells (see Figure~\ref{fig:PDF}, bottom panels). At low $r$, the two sharp maxima at 1.42 and 2.45~\AA{} correspond to the nearest-neighbor aromatic C--C bond and the second-neighbor C$\cdots$C distance across the aromatic rings, respectively. In the intermediate-$r$ region, maxima at 3.77 and 5.25~\AA{} for COF-682 and at 3.77 and 5.02~\AA{} for COF-612 arise from longer intra-linker vectors spanning the fused cores and the pendant aryl arms. At higher $r$, the peaks at 8.74 and 11.75~\AA{} for COF-682 encode inter-linker distances across the imine linkage while the peaks at 7.90 and 9.95~\AA{} for COF-612 receive contributions from both intra-linker and inter-linker correlations across the imine bond. Comparison of the two PDF peak assignments (see Figure~S6, as well as Tables S1 and S2) shows that the frameworks differ most clearly between 5 and 12~\AA{}. In COF-682, inter-linker correlations set in around 7.5~\AA{}. In COF-612, the same range remains dominated by intra-HBC distances, with inter-linker HBC$\cdots$HAPT correlations predominantly contributing to the PDF signal only beyond $\sim$11~\AA{}. This difference is a direct consequence of the larger nanographene linker of COF-612.

\noindent We then turned to investigating the dynamic properties of the two COFs,
beginning with COF-682 and the time evolution of the DHP$\cdots$DHP stacking along
the MD trajectory. As depicted in Figure~S7, the interacting DHP acene cores populate
two distinct $\pi$-stacking modes that differ both in their average interplanar
separation and in the reorientational ring-plane tilt angle $\theta$ between the
stacked acenes. In the first mode, the acene cores remain nearly parallel and stack
face-to-face, with a small tilt angle ($\theta \approx 10^\circ$, Figure S8) and a mean
interplanar distance of $\sim$4.85~\AA{} (Figure~\ref{fig:fig3}a, solid blue line).
In the second, the cores adopt an offset, tilted arrangement with a markedly larger
tilt angle ($\theta \approx 40^\circ$, Figure S8) and a correspondingly shorter interplanar
distance of $\sim$3.6~\AA{} (Figure~\ref{fig:fig3}a, solid red line). Both tilt angles
remain centered on these values throughout the trajectory, indicating that the two
modes coexist as persistent, well-defined stacking motifs. Although the mean stacking distances remain stable, the instantaneous distances evolve markedly, as evidenced by the widths of the standard-deviation envelopes (Figure~\ref{fig:fig3}a, shaded bands), which span $\sim$4.4--5.2~\AA{} for the face-to-face mode and $\sim$3.0--4.3~\AA{} for the tilted mode. The average face-to-face distance also varies more over time whereas the tilted mean stays comparatively flat, indicating that the face-to-face contact is the more flexible of the two stacking modes. Representative snapshots at $t = 0$, 7, 23,
and 40~ps illustrate these local rearrangements (Figure~\ref{fig:fig3}a, insets), which leave the COF connectivity intact
even as the acene cores continuously sample a range of stacking geometries. This structural variability is encoded in the
experimental PDF, broadening the spread among
the individual MD-snapshot PDFs most visibly in the $\sim$4--5.2~\AA{}
range (Figure~\ref{fig:PDF}a, gray curves).

\noindent To resolve how this behaviour influences the translational properties of the individual building units of COF-682, we monitored the root-mean-square displacement (RMSD) of the two linkers, separating the aromatic from the non-aromatic $sp^3$ carbons. For both the DHP (L1) and the TAM linker (L2), the RMSD fluctuates about equilibrium values over the trajectory (Figure S10). As shown in Figures \ref{fig:fig3}b,c the RMSD distribution for the aromatic carbon atoms exhibits larger values than that of the non-aromatic carbons, with values approximating 
$\sim$1.1 versus $\sim$0.8~\AA{} for L1 and $\sim$0.9 versus $\sim$0.6~\AA{} for L2. This result indicates that for both COF linkers the peripheral aromatic rings are more mobile than the $sp^3$ carbons that anchor the linker cores and junctions. The bounded amplitude of these displacements (below $\sim$1.2~\AA{} on average) is consistent with the well-defined PDF peaks present in the PDF experimental data.

\noindent The reorientational freedom of the aromatic rings constitutes a second key dynamical mode. It governs pore accessibility, guest diffusion, and the electronic communication between conjugated units \cite{nanoporous_flexibility,CCS}. To
quantify it, we computed the equilibrium tilt angle
of each aromatic ring relative to its reference orientation (Figure 3d). In particular, in the analysis we distinguished the fused
rings of the DHP core (L1 fused) from the pendant phenyl rings of the two linkers (L1 pendant
and L2 pendant). The fused rings of the DHP core reorient by a median of $\sim$12.1\textdegree{}, whereas the pendant aryl rings tilt over a wider range, with medians of $\sim$22.6\textdegree{} for L1 and $\sim$20.5\textdegree{} for L2. Within the DHP acene core the fused rings tilt slightly differently, the terminal rings ($\sim$14\textdegree{}) reorienting more than the inner ones ($\sim$11\textdegree{}) (Figure~S11). The rigid, $\pi$-stacked acene cores therefore anchor the framework, while the peripheral phenylene rings librate about their connecting bonds. The pendant rings are the main source of local reorientational flexibility in COF-682.

\begin{figure*}[h!]
    \centering
    \includegraphics[width=0.8\linewidth]{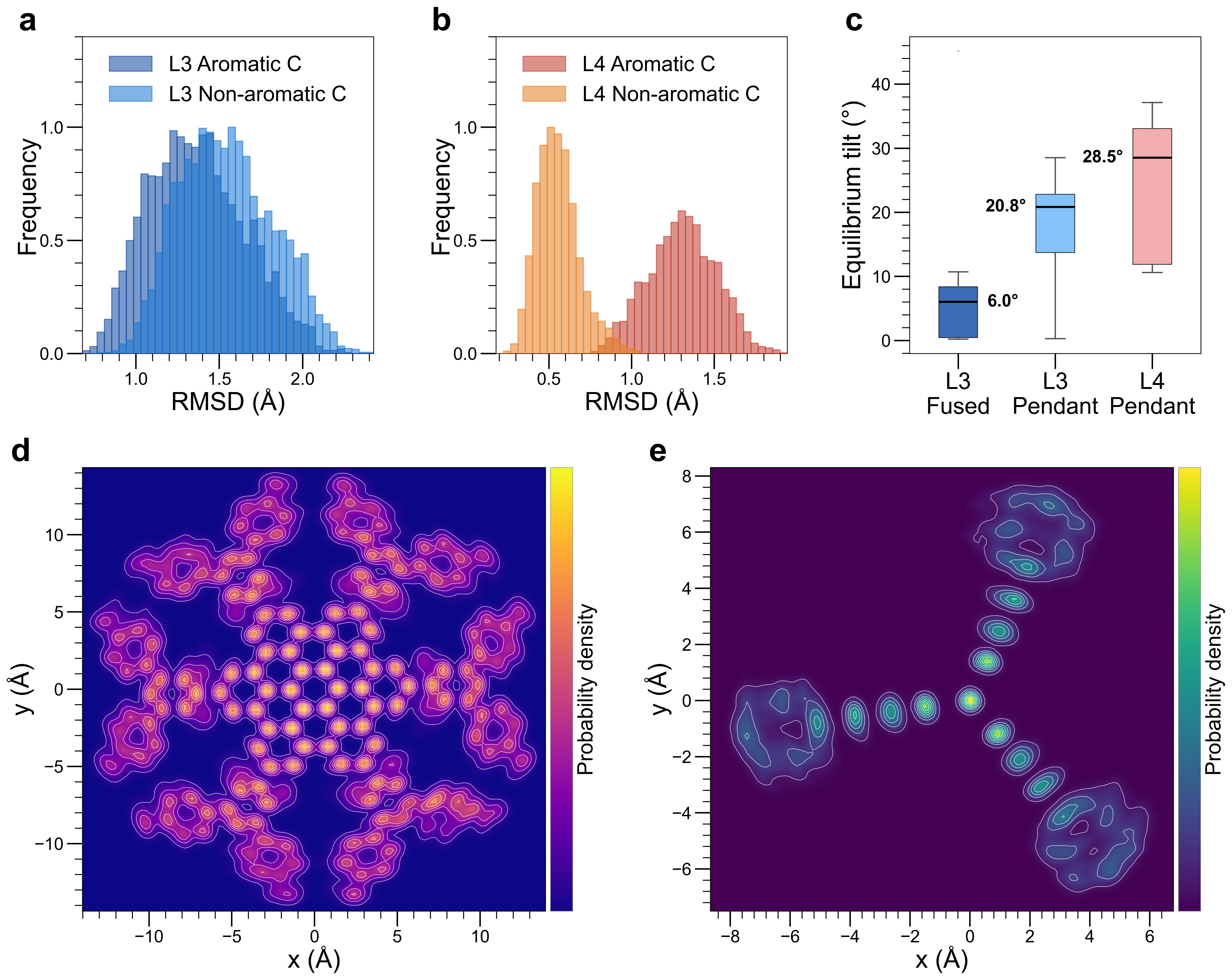}
\caption{MD-derived dynamics of COF-612. (a,b) RMSD of the
aromatic and non-aromatic carbons of linker L3 (HBC-LA$_{12}$, a) and linker L4 (HAPT, b). (c) Box plot distributions of the equilibrium tilt angles of
the fused rings of the HBC core (L3 fused) and of the pendant aryl rings (L3 pendant, L4 pendant). (d,e) Time-averaged two-dimensional positional probability densities of
representative L3 (d) and L4 (e) linkers.}
\label{fig:fig4new}
\end{figure*}

\noindent To gain insight into the real-space dynamics of the COF linkers, we computed their time-averaged positional probability densities (Figures~\ref{fig:fig3}e,f). For DHP (Figure~\ref{fig:fig3}e), the density is diffuse and ring-shaped, reflecting the mobility of the linker engaged in $\pi$ interactions, yet punctuated by sharp maxima at the positions of the pendant aromatic rings. We attribute these maxima to the libration of the pendant rings between preferred orientations, at which their atoms reside longest. The same trend is found for TAM (L2, Figure~\ref{fig:fig3}f), which lacks a fused core. As shown in Figure~\ref{fig:fig3}f, the four pendant phenylene rings of the TAM linker spread into broad densities punctuated by maxima at their preferred orientations. These probability maps locate the conformational flexibility of COF-682 directly in real space, and show it to be concentrated in the $\pi$-stacked DHP units and in the pendant rings of the DHP and TAM linkers.

\noindent We next applied the same analysis to COF-612. This COF lacks the parallel-displaced $\pi$-stacking motif of COF-682, so we examined its dynamics directly through the motion of its building units. The RMSD of the two linkers, resolved by carbon environment, fluctuates about stable values throughout the trajectory (Figure S14) while the RMSD distributions allow to differentiate the translational behaviour of the aromatic and non-aromatic carbon atoms (Figure \ref{fig:fig4new}a). For the nanographene linker HBC-LA$_{12}$ (L3), the aromatic carbons display an RMSD of $\sim$1.3~\AA{} while the non-aromatic carbon atoms display an RMSD of $\sim$1.5~\AA{} (Figures~\ref{fig:fig4new}a and S4), reversing the ordering found in COF-682. This inversion follows from the chemical identity of the non-aromatic carbons: in HBC-LA$_{12}$ these are the methoxy and imine carbons, which are located at the linker periphery, whereas in DHP they include the alkene and bridge carbons of the acene core. For the hexatopic amine HAPT (L4), the ordering matches that of COF-682, with the aromatic carbons ($\sim$1.3~\AA{}) far more mobile than the non-aromatic ones ($\sim$0.55~\AA{}), the latter corresponding to the anchored triptycene bridgeheads (Figure~\ref{fig:fig4new}b). The absolute displacements of COF-612 exceed those of COF-682 ($\sim$0.6--1.1~\AA{}). This reflects the  larger size of the HBC-based linker, whose peripheral atoms lie farther from their anchoring points, so that a given angular motion translates into a larger displacement.

\noindent The reorientational analysis provides the clearest distinction between the two COFs (Figure~\ref{fig:fig4new}c). In COF-612 the fused rings of the HBC core tilt by a median of only 6.0\textdegree{}. This is less than half the 12.1\textdegree{} median value found for the fused acene core of DHP in COF-682. The nanographene core therefore remains essentially planar and locked in orientation over the whole MD trajectory. The pendant rings, by contrast, librate over a much wider range, with medians of 20.8\textdegree{} for L3 and 28.5\textdegree{} for L4. These values are comparable to those of the pendant aromatic rings in L1 and L2 within COF-682 (22.6\textdegree{} and 20.5\textdegree{}, respectively). Consequently, the conformational flexibility of COF-612 is confined almost entirely to its peripheral rings.

\noindent The time-averaged positional probability densities give the real-space counterpart of this dynamical picture (Figures~\ref{fig:fig4new}d,e). For HBC-LA$_{12}$ (Figure~\ref{fig:fig4new}d), the fused nanographene core is resolved into a regular array of discrete, atomically sharp maxima that preserve its sixfold symmetry. These well-defined maxima persist in both the $x$--$y$ and $x$--$z$ planes (Figure~S5), in line with the conserved quasi-planarity of the core. The probability density of the nanographene core therefore exhibits far sharper colored regions than that of the acene core of COF-682, mirroring its smaller tilt (6.0\textdegree{} versus 12.1\textdegree{}). The peripheral methoxyphenyl and formylphenyl arms, by contrast, appear as broader, ring-shaped densities, reflecting the greater reorientational freedom of these pendant rings. For HAPT (Figure~\ref{fig:fig4new}e), the triptycene arms are likewise sharply defined, whereas the terminal aminophenyl rings spread into diffuse lobes due to their higher librational motion. Taken together, the RMSD, ring-tilt and density analyses show that COF-612 combines a rigid, planar nanographene core with flexible peripheral rings, in contrast to COF-682, where the flexibility extends to the $\pi$-stacked acene moiety itself.

\section*{Conclusion}

In summary, we have determined the local structure and dynamics of two three-dimensional imine-linked COFs, COF-682 and COF-612, by combining synchrotron X-ray PDF analysis with MLIP-based MD simulations. The combined PDF and MD approach developed here is widely applicable and may be extended to other imine-linked COFs as well as to COFs joined by other linkages, because it requires only that the sample retain sufficient short and intermediate range order. Our analysis shows that the local dynamics of a COF are set by a balance between the rigidity of the fused linker core, its capacity to engage in non-covalent interactions, and the mobility of the pendant aromatic groups. In COF-682, the exposed $\pi$-surface of the V-shaped dihydropentacene linker drives parallel-displaced $\pi$-stacking that samples two coexisting modes, a face-to-face and a tilted arrangement, whose interplanar distances fluctuate about 4.85 and 3.6~\AA{} at 290~K, so that flexibility extends to the acene core itself. In COF-612, by contrast, the extended nanographene core remains essentially planar and reorientational flexibility is confined almost entirely to the pendant rings. Taken together, these findings define three levers for tuning framework dynamics: choosing linkers whose fused aromatic core has a targeted size and planarity, balancing the rigidity of these aromatic cores against the libration of the pendant aryl arms, and harnessing non-covalent $\pi$-stacking between linkers to introduce controlled, local flexibility. \\
We believe that the combined PDF and MD strategy carries the characterization of COFs beyond the average periodic picture of Bragg crystallography, and that extending it from \emph{ex situ} to \emph{in situ} conditions will aid the precision engineering of reticular frameworks whose function depends on their local dynamics.

\section*{Data Availability}
Experimental and theoretical PDF data are available at
\textcolor{blue}{https://doi.org/10.5281/zenodo.21998650}

\section*{Supplementary Information}
General experimental methods and materials; COF syntheses; additional PDF and MD results. 

\section*{Acknowledgments}
F.T. acknowledges funding from the European Union’s Horizon Europe research and innovation program under the Marie Skłodowska-Curie grant agreement No. 101201413 (AQUAFRAME). Effort of S.J.L.B. and C.M. on PDF analysis, modeling and writing was supported by U.S. Department of Energy, Office of Science, Office of Basic Energy Sciences (DOE-BES) under contract No. DESC0024141. This research used beamline 28-ID-1 of the National Synchrotron Light Source II, a U.S. DOE Office of Science User Facility operated for the DOE Office of Science by Brookhaven National Laboratory under Contract No. DE-SC0012704. Claude-Opus4.8 was used for grammar and sentence structure checking during the review and editing of this manuscript.

\providecommand{\latin}[1]{#1}
\makeatletter
\providecommand{\doi}
  {\begingroup\let\do\@makeother\dospecials
  \catcode`\{=1 \catcode`\}=2 \doi@aux}
\providecommand{\doi@aux}[1]{\endgroup\texttt{#1}}
\makeatother
\providecommand*\mcitethebibliography{\thebibliography}
\csname @ifundefined\endcsname{endmcitethebibliography}
  {\let\endmcitethebibliography\endthebibliography}{}

\end{document}